\documentclass{article}
\usepackage{spconf,amsmath,graphicx}
\usepackage{cite}
\usepackage{amsmath,amssymb,amsfonts}
\usepackage{algorithmic}
\usepackage{graphicx}
\usepackage{textcomp}
\usepackage{xcolor}
\usepackage{ucs}
\usepackage[utf8x]{inputenc}
\usepackage{cite, amsfonts, amssymb, amsthm, bm, bbm, graphicx, relsize, multirow, booktabs, tikz,subfigure,soul}
\usepackage[american]{babel}
\usepackage[T1]{fontenc}
\usepackage{algorithmic, algorithm}
\usepackage[multiple]{footmisc}

\newcommand{\ds}{\displaystyle}
\newcommand{\bbeta}{\boldsymbol{\eta}}

\title{UPLINK POWER CONTROL IN CELL-FREE MASSIVE MIMO VIA DEEP LEARNING}
\name{Carmen D'Andrea$^{*}$, Alessio Zappone$^{\dagger}$, Stefano Buzzi$^{*}$ and Merouane Debbah$^{\dagger §}$ }
\address{$^{*}$DIEI, University of Cassino and Southern Latium, Cassino, Italy \\
$^{\dagger}$LANEAS group of CentraleSupelec, Gif-sur-Yvette, Paris, France  \\
$^{§}$Mathematical and Algorithmic Sciences Laboratory, France Research Center, \\ Huawei Technologies, Paris, France
}
\begin{document}
\ninept
\maketitle
\begin{abstract}
This paper focuses on the use of a deep learning approach to perform sum-rate-max and max-min power allocation in the uplink of a cell-free massive MIMO network. In particular, we train a deep neural network in order to learn the mapping between a set of input data and the optimal solution of the power allocation strategy. Numerical results show that the  presence of the pilot contamination in the cell-free massive MIMO system does not significantly affect the learning capabilities of the neural network, that gives near-optimal performance. Conversely, with the introduction of the shadowing effect in the system the performance obtained with the deep learning approach gets significantly degraded  with respect to the optimal one.
\end{abstract}
\section{Introduction}
\label{sec:intro}

The cell-free massive MIMO systems were introduced in \cite{Ngo_CellFree2016}, where one base station (BS) with a large number of antennas is replaced with a large number of low-complexity access points (APs) that serve all, or a subset of users in the system. In this scenario, the traditional concept of ``cell'' is overcome\cite{BuzziWCL2017,cell-free_downlinkpilots,Ngo_EnergyEfficiency2018} and it is more appropriate to speak about ``dynamic association rules'' in order to select the users and the APs for each communication pair in the system. 
With respect to a traditional multicell massive MIMO system with co-located arrays, cell-free systems are capable of alleviating the cell-edge problem, providing more uniform performance across users; moreover, since each user is served by multiple APs, there is also a beneficial large-scale fading diversity effect. 
In this kind of architecture, all the APs are connected via a backhaul network to a central processing unit (CPU), which, based on the association rules, sends to the APs the data symbols to be transmitted to the users in the downlink phase and receives soft estimates of the received data symbols from the APs in the uplink phase. Neither channel estimates nor beamforming vectors are propagated through the backhaul network.  One example of a practical deployment of cell-free massive MIMO could be the radio stripes \cite{Radio_Stripe_patent}. 
In this paper we consider two uplink power allocation techniques, the first one is devoted to the sum-rate maximization and the second one to the minimum-rate maximization. We propose a deep learning approach to solve these problems with a reduced computational complexity compared to the computational cost of the optimal solution. A deep artificial neural
network (ANN) is trained to learn the map between the input and the optimal power allocation strategies, and then it is used to predict the power allocation profiles for a new set input. A deep learning-based power allocation in a massive MIMO system in colocated setup was analyzed in \cite{Sanguinetti_Asilomar2018}, where the authors show that, in a scenario without shadowing effect, the performance obtained with the deep learning approach are very close to the optimal one. In this paper we consider three different scenarios, two without shadowing effect, with and without pilot contamination, and the last one with shadowing and without pilot contamination effect. In the scenarios without shadowing, we consider as input the positions of the users in the network and numerical results show the good matching in terms of rate per user of the deep learning solution and the optimal one, a similar behaviour is observed in the case of co-located massive MIMO in reference \cite{Sanguinetti_Asilomar2018}. When shadowing is considered in the system, we consider as input the coefficients containing both the path-loss and the shadowing effect and we observe that the learning capabilities of the ANN get worse with respect to the case without shadowing.

One issue with cell-free massive MIMO systems is the large size of the system in terms of access points and users to serve, which makes it more complex to perform optimal resource allocation. In this context, recently it has been observed that, thanks to the universal function approximation property of artificial neural networks (ANNs) \cite{Hornik1989}, deep learning by ANNs enables to perform radio resource allocation with a significantly lower online complexity than traditional optimization-oriented methods,  \cite{TCOM_AItutorial,ZapVTMMAG19}. In \cite{Sun2017} fully-connected ANNs are used to emulate the performance of the WMMSE power control method from \cite{LuoWMMSE}. In \cite{Liang2018} again power control by a fully-connected neural network is discussed, and it is proposed to employ the errore in the rate function as training cost function. In \cite{ZapASILOMAR2018a,ZapASILOMAR2018b} multi-cell massive MIMO systems are considered, performing power control and user-cell association. 

However, none of these previous works considers the use of ANNs for cell-free massive MIMO systems, while this appears as a relevant application given the large complexity that is incurred by optimized power control in cell-free systems. This work aims at filling this gap, developing an ANN-based uplink power control method for cell-free massive MIMO system, for the maximization of either the system sum-rate or the minimum of the users' rate. In both cases, the proposed method requires an extremely limited computational complexity, and can operate with both pilot contamination and shadowing. If no shadowing is present, the optimized power control policy is computed based only on the geographical positions of the users in the coverage area, whereas if also shadowing is present, its realizations are needed to compute improved power allocations. 

This paper is organized as follows. Next section contains the system model for the uplink cell-free massive MIMO network deployment, while Section 3 discusses the uplink power allocation optimization problems that will be solved through ANNs. Section 4 contains the description of the ANNs used to approximate the optimal uplink power allocation strategies, along with the discussion of the numerical results. Finally, conclusing remarks are given in Section 5. 

\section{System model}
We consider a square area with $K$ sigle antenna MSs and $M$ APs with $N_{\rm AP}$ antennas connected, by means of a backhaul network, to a CPU wherein data-decoding is performed. We denote as $\mathcal{K}_m$ and $\mathcal{M}_k$ the set of MSs served by the $m$-th AP and the set of APs serving the $k$-th MS, respectively.

We denote by $\mathbf{g}_{k,m}$  the $N_{\rm AP}$-dimensional vector representing the channel between the $k$-th MS and the $m$-th AP. We have $\mathbf{g}_{k,m}=\sqrt{\beta_{k,m}} \mathbf{h}_{k,m}$, with $\mathbf{h}_{k,m}$ an $N_{\rm AP}$-dimensional vector whose entries are i.i.d ${\cal CN}(0,1)$ RVs, modeling the fast fading, and $\beta_{k,m}$ a scalar coefficient given by
\begin{equation}
\beta_{k,m}= 10^{\frac{\text{PL}_{k,m}}{10}} 10^{\frac{\sigma_{\rm sh}z_{k,m}}{10}},
\label{beta_expression}
\end{equation}
where $\text{PL}_{k,m}$ represents the path loss (expressed in dB) from the $k$-th MS to the $m$-th AP, and $10^{\frac{\sigma_{\rm sh}z_{k,m}}{10}}$ represents the shadowing with standard deviation $\sigma_{\rm sh}$ \cite{Ngo_CellFree2016,BuzziWCL2017}.

At the receiver, channel estimation is performed by the linear minimum-mean-square-error (LMMSE) procedure in \cite{Ngo_CellFree2016}, denoting by $\tau_p$ the length (in time-frequency samples) of the uplink training phase, the $m$-th AP forms a LMMSE estimate of $\left\{\mathbf{g}_{k,m}\right\}_{k\in \mathcal{K}_m}$ based on the $N_{\rm AP}$-dimensional statistics $\widehat{\mathbf{y}}_{k,m}= \sqrt{\eta_k}\mathbf{g}_{k,m} + \ds \sum_{\substack{i=1 \\ i\neq k}}^K {\sqrt{\eta_i}\mathbf{g}_{i,m}\boldsymbol{\phi}_i^H \boldsymbol{\phi}_k} + \widetilde{\mathbf{w}}_{k,m}$, where ${\eta}_k$ is the power employed by the $k$-th user during the training phase, $\boldsymbol{\phi}_k$ the $\tau_p$-dimensional column pilot sequence sent by the $k$-th user, $\|\boldsymbol{\phi}_k\|^2=1$, $ \forall \, k$ and $\widetilde{\mathbf{w}}_{k,m}$ a $N_{\rm AP}$-dimensional vector with i.i.d. ${\cal CN}(0, \sigma^2_w)$ entries containing the thermal noise contribution. The LMMSE channel estimate of the channel $\mathbf{g}_{k,a}$ can be written as \cite{Ngo_CellFree2016}

\begin{equation}
\hat{\mathbf{g}}_{k,m}=\frac{\sqrt{\eta_k}\beta_{k,m}}{\ds \sum_{i=1}^K \eta_i \beta_{i,m} \left|\boldsymbol{\phi}_i^H \boldsymbol{\phi}_k\right|^2 +\sigma^2_w} \widehat{\mathbf{y}}_{k,m} = \alpha_{k,m} \widehat{\mathbf{y}}_{k,m} \; .
\end{equation}

After the channel estimation phase, the uplink data transmission phase starts. Since users do not perform channel estimation, they just send their data symbols without any channel-dependent phase offset and the generic $m$-th AP decodes only the data transmitted by users in ${\cal K}_m$ \cite{BuzziWCL2017}. After some algebraic manipulations, the soft estimates for the data sent by the $k$-th user at the CPU can be written as 
\begin{equation}
\begin{array}{llll}
\widehat{x}_k^{\rm UL} &= \ds \sum_{m\in{\cal M}_k} \ds \sqrt{\eta_{k}^{\rm UL}} \widehat{\mathbf{g}}_{k,m}^H  \mathbf{g}_{k,m} {x}_k^{\rm UL} \\ & +
\ds \sum_{\substack{j=1 \\ j\neq k}}^K \ds \sum_{m \in \mathcal{M}_k} \sqrt{\eta_{j}^{\rm UL}}\widehat{\mathbf{g}}_{k,m}^H   \mathbf{g}_{j,m} {x}_j^{\rm UL} +  
 \ds \sum_{m\in{\cal M}_k} {\widehat{\mathbf{g}}_{k,m}^H \mathbf{w}_m  } .
\end{array}
\label{Est_UL}
\end{equation}
with ${\eta_{k}^{\rm UL}}$ and ${x}^{\rm UL}_k$  representing the uplink transmit power and the data symbol of the $k$-th user, respectively, and $\mathbf{w}_m \sim {\cal CN}(\mathbf{0}, \sigma^2_w \mathbf{I}_{N_{\rm AP}} )$ is the $N_{\rm AP}$-dimensional noise vector.
\begin{figure*}
\begin{equation}
\mathcal{R}_{k}^{\rm UL}= \ds\frac{\tau_{u}}{\tau_c}  W \log_2 \left( 1+ \frac{ \eta_{k}^{\rm UL} \left( \ds \sum_{m\in{\cal M}_k} {\ds  \gamma_{k,m}} \right)^2}{
\ds \sum_{j=1}^K \eta_{j}^{\rm UL}  \sum_{m\in{\cal M}_k} \beta_{j,m} \gamma_{k,m} + \ds \sum_{\substack{j=1 \\ j \neq k}}^K \eta_{j}^{\rm UL}\left( \ds \sum_{m \in {\cal M}_k}  \gamma_{k,m} \frac{\beta_{j,m}}{\beta_{k,m}}\right)^2  \left|\boldsymbol{\phi}_j^H \boldsymbol{\phi}_k \right|^2 + \sigma^2_w \!\!\!\!\sum_{m\in{\cal M}_k} {\!\! \gamma_{k,m}}
} \right) 
\label{eq:SE_UL}
\end{equation}
\hrulefill
\end{figure*}

\section{Power allocation strategies} \label{Power_allocation_section}
As performance measures for the power control strategies, we consider the lower bound expressions of the uplink achievable rate. The expression, computed through the use-and-then-forget (UatF) bounding techniques in \cite{marzetta2016fundamentals,bjornson2017massive} is reported in Eq. \eqref{eq:SE_UL} at the top of next page, where $W$ is the system bandwidth, $\tau_{u} $ is  the length (in samples) of the uplink data transmission phases in each coherence interval, $\tau_c$ is the length (in samples) of the coherence interval and $\gamma_{k,m}=\sqrt{\eta_k}N_{\rm AP}\alpha_{k,m} \beta_{k,m}$. 
The details of the derivation are here omitted due to the lack of space.

We consider that the transmit powers are allocated in order to maximize either the system sum-rate or the minimum of the users' rate, two problems that are respectively formulated as:
\begin{subequations}\label{Prob:SumRateUL}
\begin{align}
&\ds\max_{\bbeta^{\rm UL}}\;\sum_{k=1}^K\mathcal{R}_{k}^{\rm UL}\left(\bbeta^{\rm UL}\right)\label{Prob:aSumRateUL}\\
&\;\textrm{s.t.}\; 0\leq \eta_{k}^{\rm UL}\leq P_{{\rm max},k}^{\rm UL} \,\forall\; k=1,\ldots,K\;,\label{Prob:bSumRateUL}
\end{align}
\end{subequations}
and
\begin{subequations}\label{Prob:MinRateUL}
\begin{align}
&\ds\max_{\bbeta^{\rm UL}}\;\min_{1,\ldots,K}\mathcal{R}_{k}^{\rm UL}\left(\bbeta^{\rm UL}\right)\label{Prob:aMinRateAppUL}\\
&\;\textrm{s.t.}\; 0\leq \eta_{k}^{\rm UL}\leq P_{{\rm max},k}^{\rm UL}\,\forall\; k=1,\ldots,K\; ,\label{Prob:bMinRateAppUL}
\end{align}
\end{subequations}
where $\bbeta^{\rm UL}$ is the $K$-dimensional vector collecting the uplink transmit powers of all MSs and  
$P_{{\rm max},k}^{\rm UL}$ is the maximum transmit power of the $k$-th user. Problems \eqref{Prob:SumRateUL} and \eqref{Prob:MinRateUL} have non-concave objective functions and a large number of optimization variables, which makes their solution challenging by traditional optimization theory methods. Recently, the framework of successive lower-bound maximization merged with alternating optimization has been proposed to tackle problems of the form of \eqref{Prob:SumRateUL} and \eqref{Prob:MinRateUL}, \cite{BuzziZappone_PIMRC2017,AloBuZapTGCN2019,RazaviyaynSIAM}. Nevertheless, given the large amount of variables to optimize, it is of interest to develop alternating methods that lend themselves to online implementation. This can be accomplished by deep learning tools, as described in the coming section. 

\section{Power allocation via deep learning and experimental results}
Leveraging the universal function approximation property of ANNs \cite{hornik1989multilayer,goodfellow2016deep}, it is possible to solve Problems \eqref{Prob:SumRateUL} and \eqref{Prob:MinRateUL} by deep learning. Specifically, an ANN can be used to learn the unknown map between the desired power control policy $\bbeta^{\rm UL, *}$ and the generic $L$-dimensional input of the ANN, say $\mathbf{x}$. In this work, the input $\mathbf{x}$ has been taken to be either the users' positions, or the shadowing coefficients. 

Particularly, we use a feedforward ANN with fully-connected layers, and consisting of a $L$-dimensional input layer, $N_L$ hidden layers, and a $K$-dimensional output layer yielding an estimate $\widehat{\bbeta}^{\rm UL}=\left[ \widehat{\eta}_{1}^{\rm UL},\ldots, \widehat{\eta}_{K}^{\rm UL} \right]$ of the optimal power allocation vector $\bbeta^{\rm UL, *}$. In order to train such an ANN, a training set containing $N_{ \rm T}$ multiple samples $ \lbrace \mathbf{x}(n), \bbeta^{\rm UL, *}(n), \; n=1,\ldots, N_{\rm T} \rbrace$ has been generated, where $\bbeta^{\rm UL, *}(n)$ corresponds to the power allocation obtained by the optimization methods from \cite{BuzziZappone_PIMRC2017,AloBuZapTGCN2019}, for the training input $\mathbf{x}(n)$. It should be stressed that both the generation of the training set and the execution of the training algorithm can be executed \emph{offline} and sporadically, i.e. at a longer time-scale than that at which the system input $\mathbf{x}$ varies. Indeed, after a training phase has been completed, the trained ANN can be used to infer the power allocation to be used when new system inputs $\mathbf{x}$ occur. In other words, the only operation that needs to be performed \emph{online}, is a forward propagation of the trained ANN, when a new input $\mathbf{x}$ occurs. This has a negligible complexity, especially in comparison with having to solve Problem \eqref{Prob:SumRateUL} or Problem \eqref{Prob:MinRateUL} by traditional methods every time $\mathbf{x}$ changes. 

The rest of this section provides more details on the adopted training procedure and analyzes the performance of the proposed ANN-based method by numerical simulation.


\subsection{Experimental results} In our simulation setup, we consider a communication bandwidth of $W = 20$ MHz centered over the carrier frequency $f_0=1.9$ GHz. The antenna height at the AP is $15$ m and at the MS is $1.65$ m. The additive thermal noise is assumed to have a power spectral density of $-174$ dBm/Hz, while the front-end receiver at the AP and at the MS is assumed to have a noise figure of $9$ dB and a square area of $500 \times 500$ (square meters) is considered. In order to emulate an infinite area and to avoid boundary effects, the square area is wrapped around \cite{Ngo_CellFree2016,BuzziWCL2017}. We assume $M=30$, $K=5$ and a pure cell-free approach, i.e., $\mathcal{K}_m=\{1,\ldots,K\} \, \forall m=1,\ldots,M$ and $\mathcal{M}_k=\{1,\ldots,M\} \, \forall k=1,\ldots,K$. 
We assume the length of the pilot sequences $\tau_p=8$, the length of the uplink data transmission phase is $\tau_{u}=\frac{\tau_c-\tau_p}{2}$, with $\tau_c=200$ samples as in \cite{Ngo_CellFree2016}. The uplink transmit power during the channel estimation is $\eta_k=\tau_p p_k$, with $p_k=100$ mW, $\forall k=1,\ldots,K$. For the power control strategies, we assume $P_{{\rm max},k}^{\rm UL}=100$ mW, $\forall \; k=1,\ldots,K$.
We assume that the APs are randomly located in the square area. The ANNs were trained based on a dataset of $N_{ \rm T}=1990000$ samples of independent realizations of the MSs' positions, and optimal power allocations $\bbeta^{\rm UL}$ obtained by solving Problems \eqref{Prob:SumRateUL} and \eqref{Prob:MinRateUL} as briefly reported in Section \ref{Power_allocation_section}. Particularly, 90\% percent of the samples was used for training and 10\% for validation. Other 10000 samples formed the test
dataset, which is independent from the training dataset. The ADAM training algorithm with Nesterov’s momentum has been employed for training \cite{sutskever2013importance,dozat2016incorporating}, and with the relative mean square error (MSE) as loss function. The ANNs were trained in two steps: in the former we use an initial learning rate of 0.002 setting the number of training epochs to 20, in the latter we start from the weights and biases of the neural network obtained at the end of the first training step and we use an initial learning rate of 0.001 setting the number of training epochs to 20. In both the training steps we use a batch size of 128. The training algorithm has been implemented using the open source python library Keras. 

\begin{table}[!t]
\caption{Layout of the ANN1. The trainable parameters are 46661}
\label{ANN_1}
\centering
\begin{tabular}{l|l|l|l}
 & Size & Parameters & Activation function \\ \hline
Input & 10 &  & - \\
Layer 1 (Dense) & 256 & 2816 & elu \\
Layer 2 (Dense) & 128 & 32896  & relu \\
Layer 3 (Dense) & 64 & 8256 & relu \\
Layer 4 (Dense) & 32 & 2080  & relu \\
Layer 5 (Dense) & 16 & 528 & relu \\
Layer 6 (Dense) & 5 & 85 & linear 
\end{tabular}
\end{table}

\begin{table}[!t]
\caption{Layout of the ANN2. The trainable parameters are 180805}
\label{ANN_2}
\centering
\begin{tabular}{l|l|l|l}
 & Size & Parameters & Activation function \\ \hline
Input & 10 &  & - \\
Layer 1 (Dense) & 512 & 5632 & elu \\
Layer 2 (Dense) & 256 & 131328 & relu \\
Layer 3 (Dense) & 128 & 32896  & relu \\
Layer 4 (Dense) & 64 & 8256 & relu \\
Layer 5 (Dense) & 32 & 2080  & relu \\
Layer 6 (Dense) & 16 & 528 & relu \\
Layer 7 (Dense) & 5 & 85 & linear 
\end{tabular}
\end{table}

\begin{table}[!t]
\caption{Layout of ANN3. The trainable parameters are 252485}
\label{ANN_3}
\centering
\begin{tabular}{l|l|l|l}
 & Size & Parameters & Activation function \\ \hline
Input & 150 &  & - \\
Layer 1 (Dense) & 512 & 77312 & elu \\
Layer 2 (Dense) & 256 & 131328 & relu \\
Layer 3 (Dense) & 128 & 32896  & relu \\
Layer 4 (Dense) & 64 & 8256 & relu \\
Layer 5 (Dense) & 32 & 2080  & relu \\
Layer 6 (Dense) & 16 & 528 & relu \\
Layer 7 (Dense) & 5 & 85 & linear 
\end{tabular}
\end{table}

\begin{table*}[h]
\caption{The MSE obtained over the training/validation sets.}
\label{Tr_Val_MSE}
\begin{tabular}{p{1cm}|p{2cm}|p{2cm}|p{2cm}|p{2cm}|p{2cm}|p{2cm}|}
\cline{2-7}
 & S1-SR: Tr/Val& S1-MR: Tr/Val&  S2-SR: Tr/Val& S2-MR:  Tr/Val& S3-SR: Tr/Val&  S3-MR: Tr/Val\\ \hline
\multicolumn{1}{|l|}{Epoch 1} & 0.0425/0.0346 & 0.0625/0.0564 & 0.0558/0.0470 & 0.0606/0.0478 & 0.0803/0.0732 & 0.0723/0.0630  \\ \hline
\multicolumn{1}{|l|}{Epoch 5} & 0.0187/0.0192 & 0.0455/0.0456 & 0.0343/0.0351 & 0.0357/0.0376 & 0.0719/0.0720 &0.0447/0.0470 \\ \hline
\multicolumn{1}{|l|}{Epoch 10} & 0.0160/0.0160 & 0.0420/0.0422 & 0.0317/0.0333 & 0.0330/0.0335 & 0.0717/0.0722 & 0.0431/0.0471 \\ \hline
\multicolumn{1}{|l|}{Epoch 15} & 0.0150/0.0175 & 0.0402/0.0403  & 0.0307/0.0320 & 0.0319/0.0339 & 0.0716/0.0714  & 0.0414/0.0436 \\ \hline
\multicolumn{1}{|l|}{Epoch 20} & 0.0143/0.0154 & 0.0390/0.0396 & 0.0302/0.0312 & 0.0310/0.0324 & 0.0715/0.0718 & 0.0407/0.042 \\ \hline
\multicolumn{1}{|l|}{Epoch 25} & 0.0129/0.0133 & 0.0364/0.0370 & 0.0287/0.0289 & 0.0282/0.0286 & 0.0713/0.0717 & 0.0396/0.0401 \\ \hline
\multicolumn{1}{|l|}{Epoch 30} & 0.0126/0.0128 & 0.0359/0.0361 & 0.0284/0.0291 & 0.0279/0.0290 & 0.0713/0.0713 & 0.0393/0.0393 \\ \hline
\multicolumn{1}{|l|}{Epoch 35} & 0.0123/0.0125 & 0.0355/0.0359 & 0.0282/0.0285 & 0.0277/0.0299 & 0.0713/0.0713 & 0.0391/0.0390 \\ \hline
\multicolumn{1}{|l|}{Epoch 40} & 0.0121/0.0127 & 0.0350/0.0357 & 0.0280/0.0288 & 0.0276/0.0279 & 0.0712/0.0712 &  0.0389/0.0407 \\ \hline
\end{tabular}
\end{table*}

\begin{figure}[!t]
\centering
\includegraphics[scale=0.5]{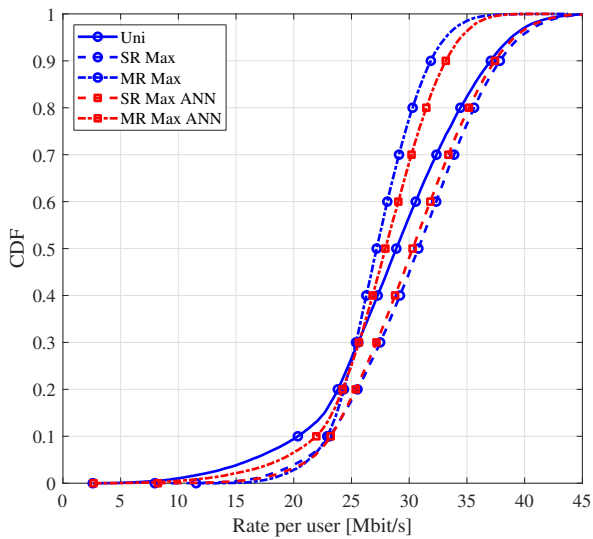}
\caption{CDFs of uplink rate per user assuming S1.}
\label{Fig:rate_UL_noshad_noPC}
\end{figure}

\begin{figure}[!t]
\centering
\includegraphics[scale=0.5]{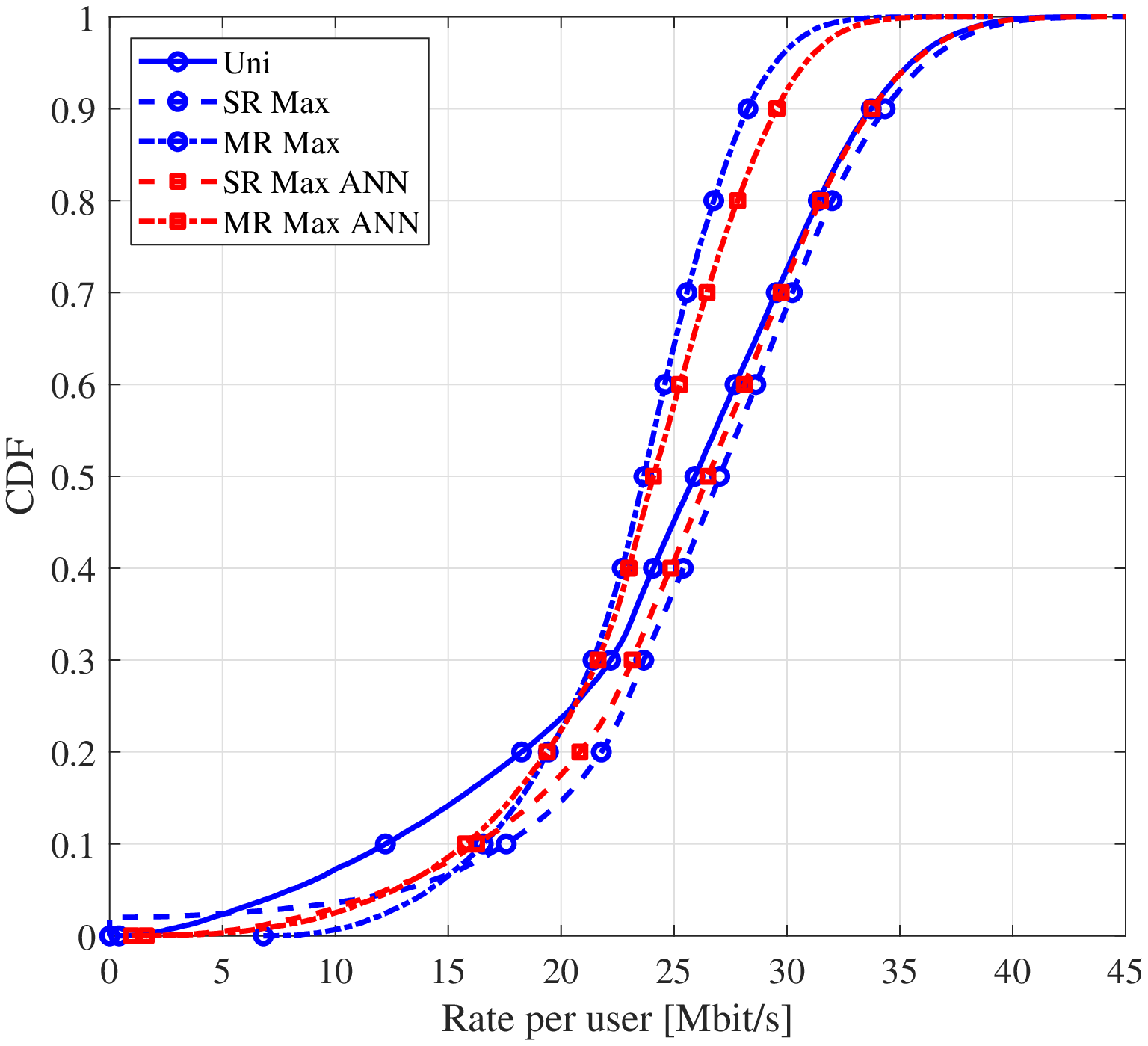}
\caption{CDFs of uplink rate per user assuming S2.}
\label{Fig:rate_UL_noshad_PC}
\end{figure}

\begin{figure}[!t]
\centering
\includegraphics[scale=0.5]{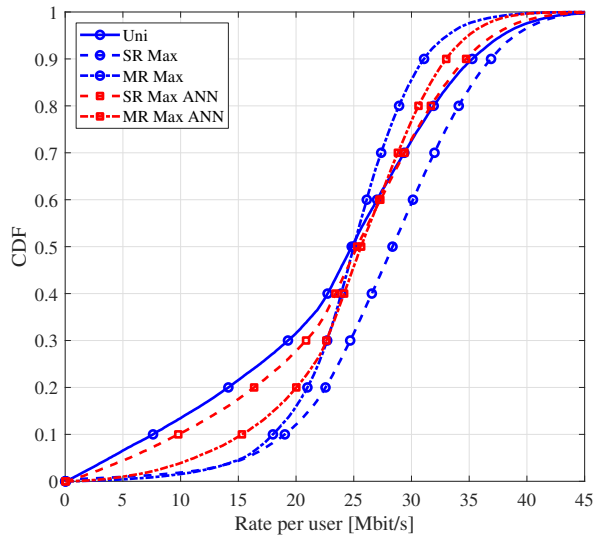}
\caption{CDFs of uplink rate per user assuming S3.}
\label{Fig:rate_UL_shad_noPC}
\end{figure}

We consider three scenarios: (i) scenario 1 (S1), without pilot contamination, i.e., the pilots sequences for all the users are orthogonal, and without shadowing, i.e., in Eq. \eqref{beta_expression} $z_{k,m}=0, \; \forall k=1,\ldots,K, m=1,\ldots,M$; (ii) scenario 2 (S2) with pilot contamination, i.e., the users' pilots are maximum-length-sequences (pseudo-noise) and without shadowing and (iii) scenario 3 (S3) without pilot contamination and with shadowing. In the following we denote by ``SR Max ANN'' and ``MR Max ANN'' the sum-rate and minimum-rate maximization obtained via deep learning, respectively, by ``SR Max'' and ``MR Max'' the optimal performance obtained solving Problems \eqref{Prob:SumRateUL} and \eqref{Prob:MinRateUL}, respectively, and by ``Uni'' the performance obtained assuming that all the users transmit with maximum power $P_{{\rm max},k}^{\rm UL}$.
In Figs. \ref{Fig:rate_UL_noshad_noPC} and \ref{Fig:rate_UL_noshad_PC} we report the performance in terms of rate per user in the cases S1 and S2, respectively. For the SR Max ANN we have used ANN1 in Table \ref{ANN_1} and for the MR Max we have used ANN2 in Table \ref{ANN_2}. In these cases the input of the neural networks are the $(x,y)$ positions of the users in the network, we can note that only this information is used to obtain in output the estimation of the optimal power allocation with the two strategies. We can note that the presence of the pilot contamination in the system does not change the learning capability of the neural networks. In Fig. \ref{Fig:rate_UL_shad_noPC} we report the performance in terms of rate per user in the case of S3. In this case we use the ANN3 in Table \ref{ANN_3} for both the SR Max ANN and MR Max ANN. In this case, in order to add information about the shadowing, the input of the network are the $\beta_{k,m}, \forall k=1,\ldots,K, \, m=1,\ldots,M$. We can see that in this case given the high variability of the input the ANN3 is not able to approximate the optimal performance with the available dataset. 
Finally, in Table \ref{Tr_Val_MSE} we report the training MSE (Tr) and the validation MSE (Val) for all the trained ANN detailed in the paper and it is seen that the ANNs neither underfits nor overfits the training data, even though in case S3 higher errors are obtained. 

\section{Conclusion}
In this paper we proposed a deep learning framework to perform power allocation in the uplink of a cell-free massive MIMO network. We considered a system with multiple antennas at the APs, a single antenna at the users' devices, with LMMSE channel estimation and maximum ratio combining. Considering the problems of sum-rate and minimum rate maximization, we train a deep neural network in order to learn the mapping between a set of input data and the solution obtained by standard optimization theory. Numerical results reveal that the presence of pilot contamination  does not significantly affect the learning capabilities of the ANN, that exhibits near-optimal performance. Instead, shadowing effects lead to quite worse performance of the ANN-based method. Further research is this aimed at designing an ANN capable of providing satisfactory performance also in the presence of shadowing. Moreover, downlink power control is another relevant research topic for future work. 

\bibliographystyle{IEEEbib}
\bibliography{References}

\end{document}